\documentclass[12pt]{book}

\usepackage[dvips]{graphicx,color}
\usepackage{makeidx,view}
\makeauthorindex
\makeindex

\begin{document}

\BookTitle{\itshape The Universe Viewed in Gamma-Rays}
\CopyRight{\copyright 2002 by Universal Academy Press, Inc.}
%\tableofcontents
\pagenumbering{arabic}

\chapter{%   %%%%%%%%% <===== TITLE of the contribution
%%%%%%%%%%% The first letter of each word should be capital letter.
Toward Ultra Short Gamma Ray Burst Ground Based Detection, SGARFACE status}

\author{%
Stephan LeBohec,  Frank Krennrich \& Gary Sleege\\%%% <== First author and second author. This case, same affiliation.
{\it Department of Physics and Astronomy,ISU Ames, IA, 50011, USA }
}
% Please note:
% One \AuthorContents{} is necessary
% for EACH CONTRIBUTION, for the contents page and
% One \AuthorIndex{} is necessary
% for EACH AUTHOR, for the index.
%
\AuthorContents{S.\ LeBohec and F.\ Krennrich} %%%%%%% <=== It is the data for CONTENTS. Please enter all author's name that should be initialized.

\AuthorIndex{LeBohec}{S.}%%%%%%% <=== It is the data for AUTHOR INDEX. Please enter a author's name that should be initialized.
\AuthorIndex{Krennrich}{F.} %%%%%%% <=== It is the data for AUTHOR INDEX. Please enter a author's name that should be initialized.

\section*{Abstract}

We present the status and motivation of the  Short GAmma Ray Front
Air Cherenkov Experiment (SGARFACE) which will be operated  
parallel to standard Very
High Energy $\gamma$-ray observations using the Whipple 10m
telescope.  SGARFACE is sensitive to $\rm 100MeV$-$\rm 10GeV$
$\gamma$-ray bursts with durations ranging from $\rm 100ns$ to
$\rm 100\mu s$ providing a fluence sensitivity as low as few $\rm
10^{-9} erg\cdot cm^{-2}$. Preliminary data taking started in 
November 2002.

\section{Introduction}
Gamma-ray bursts observed with space-based detectors span a
wide range of durations limited by the detector
integration time to time scales $\rm t \ge 1ms$. Satellite-based
detectors are not efficiently triggered by pulses with time scales shorter 
than $\rm \sim 1ms$ because their effective collection area is too small to 
allow for faster time sampling of signals with typical fluences. Nevertheless,
 among the detected bursts, millisecond and sub-millisecond variability is 
common (Walker \& Schaefer 2000).
\begin{figure}[t]
  \begin{center}
    \includegraphics[height=17pc]{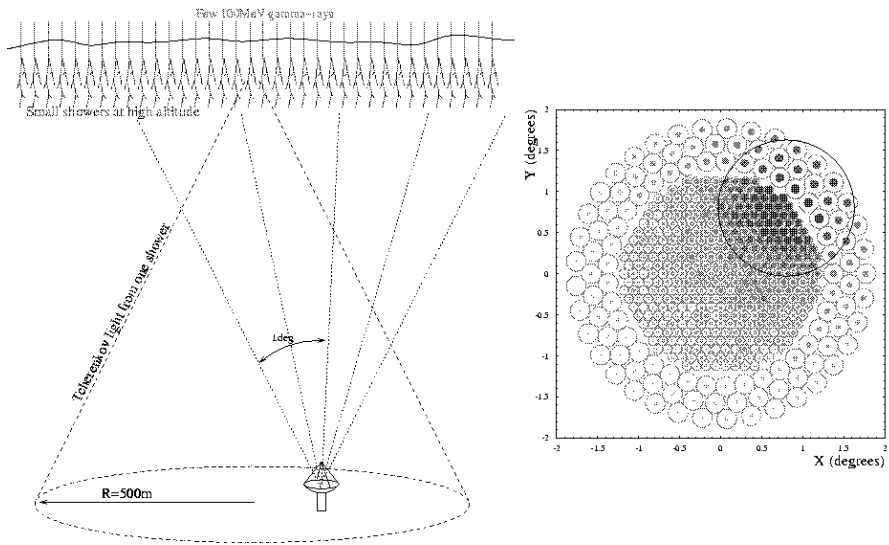}
  \end{center}
  \caption{In an imaging telescope, $\gamma$-ray bursts should appear as a Cherenkov glows.}
\label{principle}
\end{figure}

A front of low energy $\gamma$-rays entering the Earth's
atmosphere results in the development of small electromagnetic cascades 
at $\rm \sim 20km$ altitude. Each of them produces Cherenkov light spread 
over an area of $\rm \sim 500m$ in radius at ground level. The Cherenkov
light from all the showers should give a detectable glow extending
over $\rm \sim2^o$ centered on the direction of the burst 
(figure~\ref{principle}). 
The
diffuse night sky background (our sensitivity limitation) of 
collected over such a solid angle during $\rm \sim10\mu s$ is
comparable to the Cherenkov light yield from a few $\rm 100MeV$
$\gamma$-ray burst with a fluence of $\rm 10$ $\gamma$-ray $\rm
\cdot m^{-2}$. This also corresponds to the fluence sensitivity of
a space-based detector with collection areas of $\rm \sim 1 m^2$. 
Going toward shorter time scales, the noise
contamination is reduced and atmospheric Cherenkov detectors gain
in sensitivity following the square root of the burst duration.
The fluence sensitivity for a 100~ns burst of 1~GeV $\gamma$-rays
is $\rm \sim 0.1 \gamma ray \cdot m^{-2}$ (Krennrich et al.,
2000).  The timing and imaging analysis allows to remove the
dominant $\rm 100ns-10\mu s$ background due to atmospheric
scintillation produced by ultra-high-energy cosmic-ray showers.
The features of the multi-$\gamma$-ray front initiated shower
images are unique. Both the light distribution and the time
structure of the Cherenkov light image are centered on the
direction of the burst and should be symmetric. We are not aware
of any phenomenon capable of producing fake signals and we expect
the technique to allow background free detection of bursts.

SGARFACE is the first ground-based experiment using the imaging and 
time sampling to search for ultra-short bursts of $\gamma$-rays. The 
Whipple observatory 10~m $\gamma$-ray telescope will be used to collect 
the Cherenkov light and form the image. After discussing the design
and status of SGARFACE, we will review some of its motivations.

\section{SGARFACE}
\begin{figure}[t]
  \begin{center}
    \includegraphics[height=11pc]{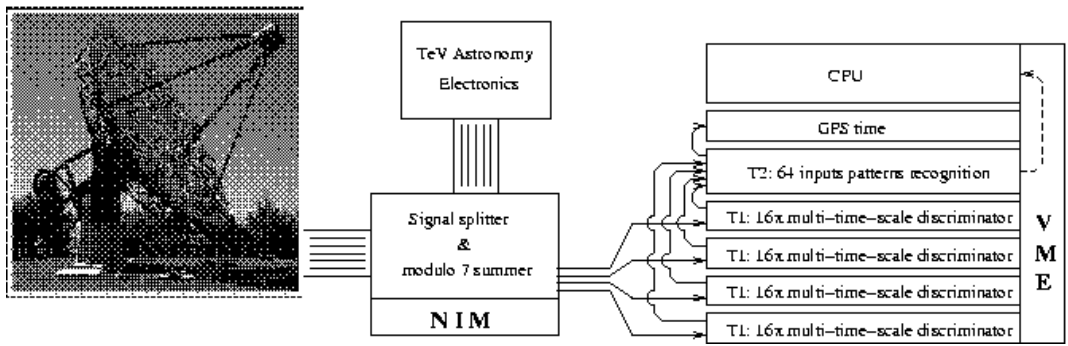}
  \end{center}
  \caption{General scheme of the SGARFACE electronics.}
\label{experiment}
\end{figure}

The photomultiplier signals coming from the Whipple 10~m (Finley
et al. 2001) Imaging Atmospheric Cherenkov Telescope are
duplicated before they reach the standard electronic system used for TeV
observations. This allows to carry out a search for bursts at the
same time the telescope is used for VHE-astronomy.
 The signals (see figure~\ref{experiment}) are sent to the Trigger~I modules
 for digitization and subsequent real time-analysis in
 Field Programmable Gate Arrays (FPGAs).
Each time the FPGAs detects a potentially interesting pulse, they send a 
signal to a coincidence unit (Trigger~II module).
\subsection{Splitter summer modules}
Since the glow produced by a $\gamma$-ray burst is quite extended
(~1deg.), it is possible to sacrifice the high angular resolution
provided by the Whipple 10m telescope camera ($\rm \sim 0.13^o$) to cost
effectiveness. The splitter summer has been designed to produce
the analog sum of seven neighboring photomultiplier signals
providing an effective angular resolution of $\rm \sim 0.4^o$. A
passive splitter preserves the bandwidth of the standard Whipple
10m electronics and reduce their amplitude by about $\rm 10\%$. Each
NIM module consists of five boards. Each board takes the inputs of
seven signals and provides 1 output channel for SGARFACE. A total
of 55 boards are necessary for the complete operation of SGARFACE
on the Whipple 10m telescope.
\subsection{Trigger-I, multi time scale discriminators}
Since the duration of the pulse to be detected is not known a
priori, the trigger decision must depend on the integrated signal
over a range of time scales. The integral of the signal over time
is derived by continuously summing the difference between the values
at the input and output of a ''first-in-first-out'' register stack 
through which the digitized signal amplitudes are driven at a 
50MHz rate.  In the SGARFACE design, the signal is integrated over 
3 contiguous time windows and a trigger signal is formed when the three 
integral values exceed a predefined threshold at the same time. This design
allows the suppression of frequent short (less than 40ns)
Cherenkov pulses produced by single particle induced cosmic-ray
showers. This logic is replicated in a cascade providing
sensitivity over time windows of width 60ns, 180ns, 540ns, 1620ns,
4860ns and 14580ns. Trigger signals from each time scale are sent
to the coincidence unit which can issue a global trigger.
The multi-time-scale discriminators are 16-channel VME-based
boards. Hence, 4 Trigger-I modules are sufficient for the entire
experiment. The digital signal processing is done by a
re-programmable Xilinx FPGA for each channel. After a trigger
occurs the local computer can read out the data present
in the multi-time-scale discriminator logic.
\subsection{Trigger-II, topological trigger}
The Trigger-II is designed to take 64 asynchronous inputs. Only
one such module is necessary for the operation of SGARFACE. It can
be made sensitive to 64 overlapping sectors in order to reduce the
accidentals rate. For each sector, the number of inputs in a high
state is calculated and compared to a multiplicity threshold. When
any of the 64 sectors reach the trigger conditions a global trigger
signal is issued. This signal is used to freeze the trigger-I
modules and hold the pulse information. It is also used to
generate an interrupt to notify the local computer that new data
is available to be read. The design of the Trigger-II module is
very similar to the Trigger-I. The logic is implemented in a
single Xilinx FPGA chip.
\subsection{Present status}
\begin{figure}[t	]
  \begin{center}	
	\includegraphics[height=11pc]{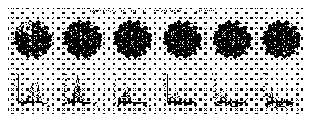}
  \end{center}	
  \caption{A cosmic ray shower seen on the event display. 
The integration time-scale ranges from left to right from 60ns to 
15$\rm \mu s$. The pulse recorded by each channel is chown in the lower 
part smoothed to the time scale corresponding to the image. The 
vertical scale is arbitrary.}
\label{event}
\end{figure}

As of November 2002, the complete system of 11 signal splitter
modules is installed and working as well as three Trigger-I boards  
 and the trigger-II module. While final tunings are made on the boards, 
various data taking strategies are being explored and we expect to start 
taking data continuously in January 2003. The first data taken toward the 
sky were encouraging and figure \ref{event} shows a cosmic ray event. From 
left to right the image acquired on time scales ranging from $\rm 60ns$ to 
$\rm 15\mu s$ are displayed and the traces under each image show the pulse 
profile from each channel smoothed to the appropriate time scale.
\begin{figure}[t]
  \begin{center}
    \includegraphics[height=16pc]{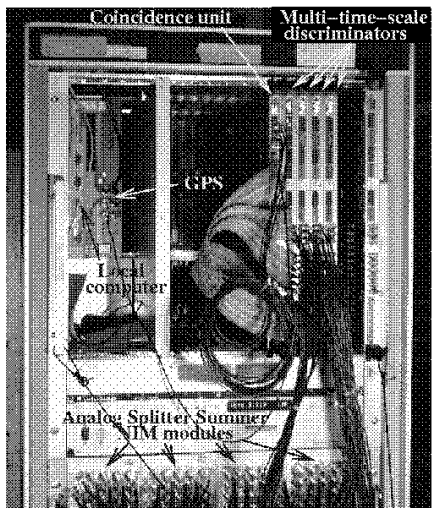}
  \end{center}
  \caption{The SGARFACE electronics installed in the Whipple 10m telescope electronics room.}
\label{electron}
\end{figure}
\section{Science perspective}
Searching for $\gamma$-ray bursts ultra-short durations has not
been explored yet with good sensitivity(Halzen et al. 1991).
Sub-millisecond structures in Gamma-Ray Bursts have been
reported by various authors. Bursts of
$\gamma$-ray with durations of less than $\rm 100\mu s$ have also been
suggested as a possible signature of the existence of primordial
black holes. In a quite different domain, some pulsars show giant radio 
pulses with durations of few microseconds which could have $\gamma$-ray 
counterparts.
\subsection{Primordial Black Holes}
%Primordial black-holes as introduced by Zeldovich in 1966 would
%result from the collapse of horizon scale density fluctuations.
%The mass function of primordial black holes depends on the
%primordial density fluctuation which are not well known and prone
%to large theoretical uncertainties (MacGibbon and Carr 1991).

%In 1974, it was shown by S.Hawking (1974) that black-holes have a
%a black body temperature inversely proportional to their mass. The
%radiated power scales as the inverse fourth power of the mass.
%Therefore, the black hole loses mass by radiation, also known as
%Hawking-radiation, at an increasingly higher rate. 
The process of
black hole evaporation leads to a violent explosion accompanied by
$\gamma$-ray emission. The time scale and details of this
explosion are mostly unknown and depend on particle physics.
However, the total energy emitted is largely model independent
$\rm \sim5\times10^{34}ergs$ and a large fraction of that likely
dissipated into  $\gamma$-rays. Short explosions of primordial
black-holes could be detected with SGARFACE as far as 250~pc from
Earth. Depending on the time scale, SGARFACE individual burst
fluence sensitivity will be up to 100 times better than that of space
based-detectors like GLAST compensating for the smaller field of view and duty cycle. 
%This is compensated by the much larger
%duty cycle and solid angle covered by space detectors. This makes
%the two types of experiments complementary and it should allow to
%probe for the existence of primordial black-holes at levels two
%orders of magnitude lower than present upper limits independently
%from the unknown particle physics at the highest energies.
\subsection{Pulsar giant pulses}
Giant radio Pulses from the Crab pulsar reach up to 2000 times the
fluence of the average single pulse. It was recently shown
(Sallmen et al. 1999) that giant radio pulse from the Crab pulsar
have typical durations of a few microseconds and less. Searches
for counterparts of giant pulses in optical, X-ray and low energy
$\gamma$-ray (Patt 1999) proved unsuccessful.  EGRET was not
sensitive to individual pulses with 2000 times the average pulsed
gamma-ray fluence. Rare microsecond bursts correlated with short
giant radio pulses could not be excluded (Ramanamurthy 1998;
Ramanamurthy 1995). SGARFACE sensitivity is such that a 100ns to $\rm 10\mu s$
 burst would be detectable
 at $\rm E> 0.25GeV$ gamma-rays if giant pulses were to occur at $\gamma$-ray
energies and exceed the average $\gamma$-ray fluence of a single
pulse by a factor of 1000 or more. SGARFACE clearly matches the
observed time scales of giant pulses and the  relative fluence
increases seen in the radio. Another good candidate for this search is 
PSR1937+21 which is is the fastest known millisecond pulsar (1.56ms) and 
exhibits giant radio pulses with amplitudes larger than 1000 times the 
average pulse amplitude.
%Giant radio radio pulses from the
%Crab Pulsar occur at a rate of 100 - 500 per hour. The amplitude
%distribution follows a power law with index of $\sim3.3$ (Lundgren
%1995). This corresponds to about 1 giant radio pulse with 1000
%times the average pulse amplitude in 20 - 30 hours of observation.
%We are planning to get a radio telescope involved to make sure
%that sufficiently strong giant pulses did in fact occur during our
%observations. Another good candidate for this search is PSR1937+21
%which is is the fastest known millisecond pulsar (1.56ms) and
%exhibits giant radio pulses with amplitudes larger than 1000 times
%the average pulse amplitude. Giant pulses occur at a hight rate
%than in the case of the Crab. A  pulse of 2000 times the average
%radio amplitude pulse occurs about every seven hours.
\section{Conclusion}
With the installation of SGARFACE at the Whipple 10m telescope, its
sensitivity will be expanded to GeV photon bursts of $\rm 0.1\mu s$ to
$\rm 100\mu s$.  SGARFACE is expected to have a fluence sensitivity of 500 times
that of EGRET for microsecond bursts of GeV photons. Data taking should start
before the end of 2002 and will be the object of an interesting search for
sub-millisecond $\gamma$-ray signal with possible implications for primordial
black holes and pulsar giant pulses.
\section{References}
\re
Walker, K.C. and Schaefer, B.E., 2000, ApJ, 537:264-269.
\re
Krennrich, F., LeBohec, S. and Weekes, T.C., 2000, ApJ, 529:506-512.
\re
F.Halzen et al., 1991, Nature, Vol. 353: 807-814.
%\re
%MacGibbon, J.H. and Carr, B.J., 1991, ApJ, 371:447-469.
\re
%Hawking, S.W., 1974, Nature 248, 30-31.
%\re
Sallmen et al., 1999, ApJ, 517: 460.
\re
Patt et al. 1999, ApJ, 522: 440.
\re
Ramanamurthy et al., 1998, ApJ, 496: 863.
\re
Ramanamurthy  et al. 1995, ApJ, 450: 791.
\re
Lundgren et al. 1995, ApJ, 453: 433.
%\newpage
%%%%%%%%%%%%%%%%%%%%%%%%%%%%%%%%%%%%%%
%       Please fill out items listed below. See %%% EXAMPLE %%%
%
%%%%%%%%%%%%%%%%%%%%%%%%%%%%%%%%%%%%%%

\chapter*{ Entry Form for the Proceedings }

\section{Title of the Paper}
%%%%%%%%%%%%%%%%% Enter the title of your paper.

{\Large\bf %
\* The Cosmic Ray Background as a Tool for Relative Calibration
of Imaging Atmospheric Cherenkov Telescopes  %%% EXAMPLE %%%
}

\section{Author(s)}

\newcounter{author}
\begin{list}%
{Author No. \arabic{author}}{\usecounter{author}}

%%%%%%%%%%%%%%%%% This item unit is just for one author.
\item %
\begin{itemize}
\item Full Name:                Stephan LeBohec %%% EXAMPLE %%%
\item First Name:               Stephan %%% EXAMPLE %%%
\item Middle Name:               %%% EXAMPLE %%%
\item Surname:                  LeBohec %%% EXAMPLE %%%
\item Initialized Name:         S.LeBohec %%% EXAMPLE %%%
\item Affiliation:              Iowa State University %%% EXAMPLE %%%
\item E-Mail:                   lebohec@iastate.edu %%% EXAMPLE %%%
\item Ship the Proceedings to:  Department of Physics and Astronomy,ISU Ames, IA, 50011, USA %%% EXAMPLE %%%
\end{itemize}

%%%%%%%%%%%%%%%%% This item unit is just for one author.
\item %
\begin{itemize}
\item Full Name:                Frank Krennrich
\item First Name:               Frank
\item Middle Name:
\item Surname:                  Krennrich
\item Initialized Name:         F.Krennrich
\item Affiliation:              Iowa State University
\item E-Mail:                   krennrich@iastate.edu
\item Ship the Proceedings to:  Department of Physics and Astronomy,ISU Ames, IA, 50011, USA
\end{itemize}

\end{list}

\endofpaper
\end{document}